\documentclass[submission,copyright,creativecommons]{eptcs}
\providecommand{\event}{ACL2 workshop 2014} 
\usepackage{tikz}
\usetikzlibrary{decorations.markings}
\usetikzlibrary{arrows}
\usepackage{amssymb}
\usepackage{amsfonts,amsmath,latexsym,stmaryrd}
\usepackage{cite}
\usepackage{theorem}
\usepackage{wrapfig}
\usepackage{alltt}
\usepackage{fancyvrb}
\usepackage{relsize}
\usepackage{multirow} 
\usepackage{hhline} 
\usepackage{listings}
\usepackage[ruled]{algorithm2e}

\usepackage{xcolor,colortbl}
\newtheorem{thm}{Theorem}[section]

 \newtheorem{definition}[thm]{Definition}
 \newtheorem{example}[thm]{Example}

\title{ACL2(ml): Machine-Learning for ACL2\thanks{This work was supported by the SICSA Industrial Proof of Concept grant ``Machine-Learning for Industrial Theorem Proving'', and EPSRC grants EP/J014222/1 and EP/K031864/1.}}
\author{J\'onathan Heras
\institute{School of Computing, University of Dundee, UK}
\email{jonathanheras@computing.dundee.ac.uk}
\and
Ekaterina Komendantskaya 
\institute{School of Computing, University of Dundee, UK}
\email{katya@computing.dundee.ac.uk}
}
\def\titlerunning{ACL2(ml): Machine-Learning for ACL2}
\def\authorrunning{J. Heras \& E. Komendantskaya}
\begin{document}
\maketitle

\begin{abstract}
ACL2(ml) is an extension for the Emacs interface of ACL2. This tool uses machine-learning 
to help the ACL2 user during the proof-development. Namely, ACL2(ml) gives hints to the user in the form of families of similar theorems, and generates auxiliary lemmas automatically. In this paper, we present the two most recent extensions for ACL2(ml). First, ACL2(ml) can suggest now families of similar function definitions, in addition to the families of similar theorems. Second, the lemma generation tool implemented in ACL2(ml) has been improved with a method to generate preconditions using the guard mechanism of ACL2. The user of ACL2(ml) can also invoke directly the latter extension to obtain preconditions for his own conjectures. 
\end{abstract}

\section{Introduction}\label{sec:intro}

ACL2 is a theorem prover that has been successfully applied in the formalisation of industrial problems~\cite{Brock96,KMM00}, and also used for pedagogical purposes~\cite{JFP:1381480,EVF07}. Even if there is a considerable difference in the difficulty of the problems that are tackled in these two contexts, there are some common challenges that are faced by experts and novice users of ACL2 alike: re-usability of libraries, discovery of auxiliary lemmas, and comparison of similarities between theorems and definitions. In the industrial scenario, the coordination of the members of a team is also a challenge: each user has his own definitions, theorems and proof-style, making the collaborative proof-development difficult. 

These challenges are partially addressed by the searching tools already available in ACL2. ACL2 provides several alternatives to browse its documentation, and search definitions and theorems in the ACL2 books. The ACL2+books combined manual can be accessed  online~\cite{xdocacl2}, using the Emacs-based ACL2-Doc browser, or using the ACL2 \verb":DOC" command at the terminal. In addition, ACL2 books can be browsed from the community-books website~\cite{acl2books}. 

The above tools can be used to search information about an ACL2 topic, access the documentation of a concrete theorem or definition, or search a keyword in the ACL2 books. Hence, these tools are useful if the user knows what he is searching for.
When the user fails to define searching parameters exactly, it may be helpful to use statistical machine-learning to detect and identify relevant patterns.
For example,  consider the following scenarios: 
 \begin{enumerate}
 \item a user in the middle of a proof does not know how to proceed, and wishes to see some similar development to extrapolate it to his particular case; 
\item  a user has the intuition that he is proving/defining something similar (or exactly the same) to one of his previous developments, but does not remember exactly  where and how he did it; and, 
\item a user in a team-development wants to know if any other member of the team has solved his current (or a similar) problem. 
 \end{enumerate}
 In these three situations, the current searching mechanisms of ACL2 are not useful because the user does  not have exact parameters for search; 
 however, it would be extremely helpful to have a tool that could automatically detect patterns across ACL2 books; capturing higher-level intuitions the user may have.

ACL2(ml)~\cite{lpar13,acl2ml} -- a machine-learning extension for the Emacs interface of ACL2 -- was created with that particular aim. ACL2(ml) uses statistical clustering to detect families of ACL2 theorems following the same pattern; and symbolic methods to automatically generate new lemmas. In particular, ACL2(ml) features the following main functions (see also Figure~\ref{acl2mldiagram}):

\begin{itemize}
 \item The user works within the Emacs environment of ACL2, and has an option to call ACL2(ml) whenever he needs to see similar patterns.
 \item Based on the user's choice, ACL2(ml) compiles the chosen libraries, and extracts significant features from the terms of ACL2 theorems and definitions.
 \item ACL2(ml) connects to machine-learning tools, and runs a number of experiments on clustering the data for each user query. Based on the results, it chooses the most reliable patterns; thus relieving the ACL2 user of the laborious step of post-processing the statistical results.
 \item If the user chooses to see only patterns related to his current theorem, ACL2(ml) would further filter the results and show the families of related theorems to the user.  
 \item Additionally, and based on the families of similar theorems, ACL2(ml) generates lemmas that may help in the proof of a given theorem. 
 \item The related theorems and generated auxiliary lemmas are displayed in a separate window.
\end{itemize}

An overview of the initial ACL2(ml) features is given in Section~\ref{sec:acl2ml}, the details of the different techniques implemented in ACL2(ml) are given in~\cite{lpar13}. In this paper, we present the two most recent extensions of ACL2(ml): \emph{definition clustering} and a \emph{generator of preconditions}. 

The initial ACL2(ml) was focused on finding families of similar theorems; however, discovering families of similar definitions can speed-up the proof-development and avoid redundancies in the code. For instance, ACL2(ml) can discover that a newly defined function was previously defined; in that case, the user can use the existing library definition and all its background theory instead of defining it from scratch. We present the definition clustering method and its applicability to find redundancies in the ACL2 books in Section~\ref{sec:clusterdef}.

The second extension that we introduce in this paper improves the lemma generation tool implemented in ACL2(ml). The method presented in~\cite{lpar13} generates a set of conjectures that could be useful to prove a given theorem. These conjectures are automatically tested with a counterexample generator, and ACL2(ml) only suggests to the user the conjectures that are not falsified. Some of the conjectures are rejected because the necessary preconditions are missed. In Section~\ref{sec:guards}, we explain the use of ACL2 guards to find preconditions for conjectures. This technique is incorporated into the lemma generation component of ACL2(ml); additionally, the user can use this extension to find preconditions of his own conjectures. 

Finally, in Section~\ref{sec:conclusions}, we conclude the paper. ACL2(ml) and the libraries used in this paper can be downloaded from~\cite{acl2ml}.


\section{Overview of ACL2(ml)}\label{sec:acl2ml}

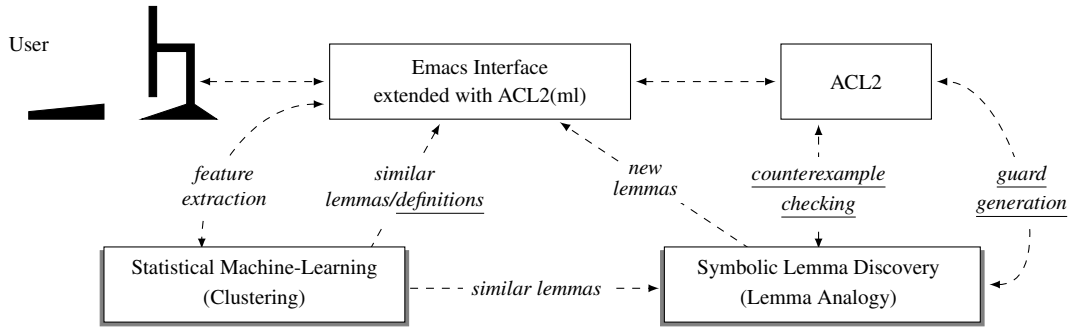
\begin{figure}
\centering 
\centering 
\footnotesize{
\begin{tikzpicture}

\draw[fill=black] (-1,2) -- (0,2) -- (0,2.2) -- (-1,2.1) --cycle;
\draw[fill=black] (.5,2) -- (1.5,2) -- (1.2,2.2) -- (1.2,3) -- (.7,3) -- (.7,3.5) -- (.6,3.5) -- (.6,2.3) -- (.7,2.3) -- (.7,2.9) -- (1.1,2.9) -- (1.1,2.2) -- cycle;

\draw[latex-latex,shorten <=2pt,shorten >=2pt,dashed] (1.2,2.5) -- (3,2.5); 
\draw (-1,3) node{{\scriptsize User}};

\draw[fill=white] (3,2) rectangle (7,3); 
\draw  (5,2.7) node{{\scriptsize Emacs Interface}}; 
\draw  (5,2.3) node{{\scriptsize extended with ACL2(ml)}}; 

\draw[latex-latex,shorten <=2pt,shorten >=2pt,dashed] (7,2.5) -- (9,2.5); 
\draw[fill=white] (9,2) rectangle (11,3); 
\draw  (10,2.5) node{{\scriptsize ACL2}}; 

\draw[latex-latex,shorten <=2pt,shorten >=2pt,dashed] (9.5,0.2) -- (9.5,2);  
\draw (9.5,1.1) node[anchor=north,fill=white]{\emph{\scriptsize{\underline{checking}}}};
\draw (9.5,1.5) node[anchor=north,fill=white]{\emph{\scriptsize{\underline{counterexample}}}};

\draw[latex-latex,shorten <=2pt,shorten >=2pt,dashed] (11.7,-0.2) to[out=0,in=0] (11,2.5);  
\draw (12.2,1.1) node[anchor=north,fill=white]{\emph{\scriptsize{\underline{generation}}}};
\draw (12.2,1.5) node[anchor=north,fill=white]{\emph{\scriptsize{\underline{guard}}}};
 
\draw[-latex,shorten <=2pt,shorten >=2pt,dashed] (9,0) -- (6,2);  
\draw (7.2,1.3) node[anchor=north,fill=white]{\emph{\scriptsize{lemmas}}};
\draw (7.2,1.5) node[anchor=north,fill=white]{\emph{\scriptsize{new}}};
\draw[fill=gray,draw=gray] (7.4,.25) rectangle (11.7,-.75);  
\draw[fill=white] (7.45,.3) rectangle (11.65,-.7); 
\draw  (9.5,0) node{{\scriptsize Symbolic Lemma Discovery}}; 
\draw (9.5,-.4) node{{\scriptsize (Lemma Analogy)}};

 \draw[latex-latex,shorten <=2pt,shorten >=2pt,dashed] (3,2.2) to[out=180,in=90] (1.3,0.2);
\draw (1.6,1.15) node[anchor=north,fill=white]{\emph{\scriptsize{extraction}}};
 \draw (1.6,1.5) node[anchor=north,fill=white]{\emph{\scriptsize{feature}}};
 
 \draw[-latex,shorten <=2pt,shorten >=2pt,dashed] (3.5,0.2) -- (4.5,2);
\draw (4,1.15) node[anchor=north,fill=white]{\emph{\scriptsize{lemmas/\underline{definitions}}}};
 \draw (4,1.5) node[anchor=north,fill=white]{\emph{\scriptsize{similar}}};

\draw[fill=gray,draw=gray] (-.05,.25) rectangle (4.05,-.75);  
\draw[fill=white] (0,.3) rectangle (4,-.7); 
\draw  (2,0) node{{\scriptsize Statistical Machine-Learning}}; 
\draw (2,-.4) node{{\scriptsize (Clustering)}};

\draw[-latex,shorten <=2pt,shorten >=2pt,dashed] (4,-0.25) -- (7.45,-0.25); 
\draw (5.75,-0.03) node[anchor=north,fill=white]{\emph{\scriptsize{similar lemmas}}};

\end{tikzpicture}}
\caption{{\scriptsize \emph{Architecture of ACL2(ml).} The Emacs interface for ACL2 extracts important features from ACL2 theorems and definitions, connects to machine-learning software, clusters the theorems there, and sends the result (families of similar theorems) to the screen. In addition, for an unproved theorem $T$ , it sends the cluster $C$ of $T$ to the lemma analogy tool, which in its turn generates auxiliary lemmas by analogy with auxiliary lemmas of the theorems in $C$. The underlined components of the diagram are the extensions presented in this paper.}}\label{acl2mldiagram} 
\end{figure}

ACL2(ml) combines \emph{statistical} and \emph{symbolic} machine-learning techniques to help the ACL2 user during the proof development (cf. Figure~\ref{acl2mldiagram}). 
In this section, we present the main functionality that the initial ACL2(ml) offers to the user.



\paragraph{Statistical machine-learning in ACL2(ml).} ACL2(ml) uses (unsupervised) clustering algorithms~\cite{Bishop} to find patterns in ACL2 theorems. Clustering algorithms divide data into $n$ groups of similar objects (called clusters), where the value of $n$ is a parameter provided by the user. In ACL2(ml), the value of $n$ is automatically computed depending on the number of objects to cluster, and using the formula provided in~\cite{lpar13}. 
Clustering methods find significant patterns among selected features. Hence, ACL2(ml) must have its own method to extract significant features from ACL2 terms.  

  We adopt the following standard terminology. We assume there is a training data set, containing some samples (or objects). \emph{Features} are given by a set of statistical parameters chosen to represent all objects in the given data set. If $n$ features are chosen, one says that object classification is conducted in an $n$-dimensional space. For this reason, most pattern-recognition tools will require that the number of selected features is limited and fixed. \emph{Feature values} are rational numbers used to instantiate the features for every given object. If an object is characterised by $n$ feature values, these $n$ values together form a \emph{feature vector} for this object. A function that assigns, to every object of the data set, a feature vector is called a \emph{feature extraction function}.

\emph{Feature extraction}~\cite{Bishop} is a research area developing methods for discovery of statistically significant features in data. ACL2(ml) implements its own feature extraction method
to collect statistical features from ACL2 terms. 
It captures the structure and dependencies of ACL2 terms by processing the whole ACL2 development.
ACL2 term trees are central to this process.

\begin{definition}[Term tree]
A variable or a constant is represented by a tree consisting of one single node, labelled by the variable or the constant itself.
A function application \verb"(f t1 ... tn)" is represented by the tree with the root node labelled by \verb"f", and its immediate subtrees given by trees representing \verb"t1", $\ldots$, \verb"tn". 
\end{definition}

The features collected from the term trees are given by a finite number of properties common to all possible trees: the term tree depth and the term arity of the function of a node. 
These parameters must be fixed at an arbitrary size. In the current standard version of ACL2(ml), we limit both parameters to $7$. Thus,  given a term $t$, we construct a $7\times 7$ table, $[t]_M$, where the rows represent term-tree depth and the columns represent term arity (where the first column is reserved for variables) -- i.e. the element $(0,j)$ of the table contains the variables of the term tree of depth $j$, and the element $(i,j)$ contains the terms of depth $j$ and arity $i-1$ (cf. Figure~\ref{tree}). 
Fixing the maximum depth and maximum arity of a node 
makes the feature extraction mechanism uniform across all ACL2 terms appearing in the libraries -- we may lose some information if pruning is needed, but the chosen size works well for most terms appearing in ACL2 libraries, and increasing the size will introduce sparsity in the feature matrices making the clustering process more difficult.

 \begin{figure}[t]
  
\begin{Verbatim}[frame=single,fontsize=\tiny,commandchars=\\\{\}]
;; \textit{Multiplication}
1 (defun mult (n m) (if (zp m) 0 (+ n (mult n (- m 1)))))
2 (defun helper-mult (n m a) (if (zp m) a (helper-mult n (- m 1) (+ n a))))
3 (defun mult-tail (n m) (helper-mult n m 0))
4 (defthm mult-mult-tail (implies (and (natp n) (natp m)) (equal (mult-tail n m) (mult n m))))
5 (defthm mult-helper-mult (implies (and (natp n) (natp m) (natp a))  (equal (helper-mult n m a) (+ a (mult n m))))

;; Exponentiation
1 (defun expt (n m) (if (zp m) 1 (* n (expt n (- m 1)))))
2 (defun helper-expt (n m a) (if (zp m) a (helper-expt n (- m 1) (* n a))))
3 (defun expt-tail (n m) (helper-expt n m 1))
4 (defthm expt-expt-tail (implies (and (natp n) (natp m)) (equal (expt-tail n m) (expt n m))))
5 ???

;; \textit{Factorial}
1 (defun fact (n) (if (zp n) 1 (* n (fact (- n 1)))))
2 (defun helper-fact (n a) (if (zp n) a (helper-fact (- n 1) (* a n))))
3 (defun fact-tail (n) (helper-fact n 1))
4 (defthm fact-fact-tail (implies (natp n) (equal (fact-tail n) (fact n))))
5 (defthm fact-helper-fact (implies (and (natp n) (natp a))  (equal (helper-fact n a) (* a (fact n))))

;; \textit{Fibonacci}
1 (defun fib (n) (if (zp n) 0 (if (equal n 1) 1 (+ (fib (- n 1)) (fib (- n 2))))))
2 (defun helper-fib (n j k) (if (zp n) j (if (equal n 1) k (helper-fib (- n 1) k (+ j k)))))
3 (defun fib-tail (n) (helper-fib n 0 1))
4 (defthm fib-fib-tail (implies (natp n) (equal (fib-tail n) (fib n)))) 
5 ???
\end{Verbatim}
\vspace{-.5cm}
 \caption{{\scriptsize \emph{ACL2 definitions and theorems.} 1: recursive arithmetic functions.  2: helpers of tail-recursive arithmetic functions.
 3: tail-recursive arithmetic functions. 4: equivalence theorems of recursive and tail-recursive functions. 5: auxiliary lemmas to prove theorems of Point 4.}}\label{fig:rec-tail}
\end{figure}  

\begin{example}\label{example1}
Consider the theorem \verb"mult-mult-tail" given in Figure~\ref{fig:rec-tail}, its term tree is depicted in Figure~\ref{tree}. This term depends on  other terms: \verb"implies", \verb"natp", \verb"equal" and so on. To extract the feature table for \verb"mult-mult-tail", see Figure~\ref{tree}, we need to know the feature values of those functions comprised in this theorem, and this is achieved using the function $[.]$. 
\end{example}

The injective function $[.] : \textnormal{ACL2 functions} \rightarrow \mathbb{Q}$ is dynamically recomputed as an ACL2 development progresses. This function is sensitive to the structure and dependencies of functions, and it assigns close values to similar functions using a \emph{recurrent clustering} process~\cite{lpar13}. Namely, given a function $f$, ACL2(ml) constructs the feature table $[f]_M$ from the definition of $f$, clusters it against the rest of the definitions of the library, and finally assigns a unique rational number, $[f]$, to the function depending on its cluster (the values of $[.]$ for the functions in the same cluster will be close, and functions in different clusters will have more distant values, see~\cite{lpar13} for the concrete formula). In order to construct the feature table $[f]_M$, we need to know the feature values of the functions comprised in $f$, and this in turn can be done by clustering their definitions, and extracting their feature values --- in the case of recursive (or mutually recursive) functions, the value for the recursive call is fixed and different to the values assigned from the clustering process. This recurrent clustering process continues up to the ACL2's built-in functions whose numerical values were pre-defined in advance. Thanks to this \emph{recurrent clustering} 
process, the 
function $[.]$ assigns 
close values to similar functions (based on the structure and function calls of their definitions), and more distant values to unrelated functions.

\begin{example}
For the functions presented in Figure~\ref{fig:rec-tail}, the function $[.]$ returned values:\\
$[$\verb"fact"$]=12.974$, $[$\verb"fib"$]=12.618$, $[$\verb"mult"$]=10.965$, $[$\verb"expt"$]=10.959$,\\
$[$\verb"helper-fact"$]=16.961$, $[$\verb"helper-fib"$]=16.431$, $[$\verb"helper-mult"$]=16.548$, $[$\verb"helper-expt"$]=16.507$,\\
$[$\verb"fact-tail"$]=18.970$, $[$\verb"fib-tail"$]=18.735$, $[$\verb"mult-tail"$]=18.699$, $[$\verb"expt-tail"$]=18.640$.\\
\end{example}

The construction of feature tables is a process that runs in the background of ACL2(ml). When the user asks ACL2(ml) to show families of similar theorem statements, ACL2(ml) initialises the recurrent clustering procedure by using a set of Emacs-Lisp scripts that communicate with  
clustering algorithms implemented in the Weka machine-learning interface~\cite{Weka}.  When the recurrent clustering cycle is completed, ACL2(ml) uses another set of scripts to cluster the statement of theorems, and post-processes the final clustering statistics 
showing  the user only the best ``guess'' of similar theorems.

\begin{example}\label{ex:multexpt}
Let us consider a library that contains theorems about the equivalence of several recursive and tail-recursive arithmetic functions including the results presented in Figure~\ref{fig:rec-tail}. In this library, we have some theorems that have been proven (theorems about multiplication and factorial functions in Figure~\ref{fig:rec-tail}) and other theorems are still unproven (exponentiation and Fibonacci theorems in Figure~\ref{fig:rec-tail}). Using this library, ACL2(ml) will show a message with clusters of similar theorems; in particular, it detects that the theorems \verb"expt-expt-tail" and \verb"mult-mult-tail" belong to the same cluster, and theorems \verb"fib-fib-tail" and \verb"fact-fact-tail" form another cluster. We can use this information to extrapolate the auxiliary lemmas that were used in the proof of multiplication and factorial to complete the proofs of exponentiation and Fibonacci respectively. 
\end{example}

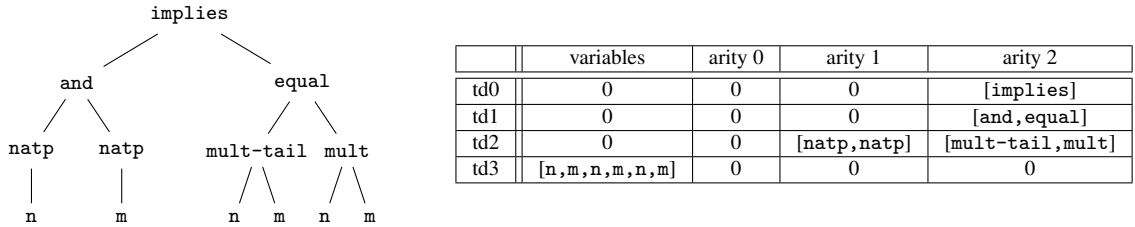
\begin{figure}
\centering
\begin{tikzpicture}

\draw (0,0) node{
\begin{tikzpicture}[level 1/.style={sibling distance=50mm},
level 2/.style={sibling distance=20mm},
level 3/.style={sibling distance=10mm},
level 5/.style={sibling distance=10mm},scale=.6,]
   \node (root) {{\scriptsize \texttt{implies}}}
               child { node {{\scriptsize \texttt{and}}}
                       child { node {{\scriptsize \texttt{natp}}}
                               child { node {{\scriptsize \texttt{n}}}}}  
                       child { node {{\scriptsize \texttt{natp}}}
                               child { node {{\scriptsize \texttt{m}}}}}  
                     }
               child { node {{\scriptsize \texttt{equal}}}
                       child { node {{\scriptsize \texttt{mult-tail}}}
                               child { node {{\scriptsize \texttt{n}}}}
                               child { node {{\scriptsize \texttt{m}}}}}
                       child { node {{\scriptsize \texttt{mult}}}
                               child { node {{\scriptsize \texttt{n}}}}
                               child { node {{\scriptsize \texttt{m}}}}
                               }
                     };
  \end{tikzpicture}};

\draw (8,0) node{
{\scriptsize 
\begin{tabular}{|c||c|c|c|c|}
\hline
& variables & arity 0 & arity 1 & arity 2 \\
\hline\hline
td0 & 0 & 0 & 0 & [\texttt{implies}] \\
\hline
td1 & 0 & 0 & 0 & [\texttt{and,equal}] \\
\hline
td2 & 0 & 0 & [\texttt{natp,natp}] & [\texttt{mult-tail,mult}] \\
\hline
td3 & [\texttt{n,m,n,m,n,m}] & 0 & 0  & 0 \\
\hline
\end{tabular}}};

\end{tikzpicture}
\caption{{\scriptsize \emph{Term tree and a fragment of the feature table of theorem \texttt{mult-mult-tail} from Figure~\ref{fig:rec-tail}.}}}\label{tree}
\end{figure}

The user may prefer the statistical hint to be related to the current theorem that he is trying to prove, as in Example~\ref{ex:multexpt}, or give information about patterns arising in the theorems of a library irrespective of the current theorem. The user may choose to data-mine only the current library, or a number of ACL2 books coming from different domains or different users. In addition, the user can configure the precision of clusters. All these options are accommodated within ACL2(ml), see~\cite{acl2ml} for a detailed description of the user interface. 

Next, we can use the statistically discovered clusters of 
theorems to generate proof hints in the form of auxiliary lemmas.

\paragraph{Lemma Analogy in ACL2(ml).} Symbolic tools for automatic discovery of lemmas, like IsaCoSy~\cite{JDB11} and IsaScheme~\cite{montano2012}, synthesise new terms using different kinds of heuristics. These tools generate candidate conjectures which are then filtered through a counterexample checker. These methods have limits: they can be slow on large inputs due to the increase in the search space, and rely on having access to good counterexample finders for filtering of candidate conjectures.

The approach followed in ACL2(ml) to synthesise new terms relies on the statistically discovered clusters, and uses term trees clustered together to synthesise new
theorems. Namely, in order to reduce the search space, ACL2(ml) feeds the output of the clustering algorithms to a \emph{mutation} tool. We adopt the following terminology. A \emph{target theorem} (TT) is a theorem currently being attempted in ACL2, but requiring user's intervention in the form of auxiliary lemmas. A \emph{source theorem} (ST) is a theorem which has been suggested as similar to the TT by the statistical ACL2(ml). A \emph{source lemma} (SL) is a lemma required for proving the ST (the SLs are obtained inspecting the event summary printed by ACL2 after proving the ST). The symbolic side of ACL2(ml) groups (potentially multiple) statistical suggestions into ST and SL pairs, each being evaluated in turn. The process then outputs some \emph{target lemmas} (TLs) -- these lemmas are analogical to some SL, and not falsified by counterexample checking.

\begin{figure}
\centering
\includegraphics[scale=.4]{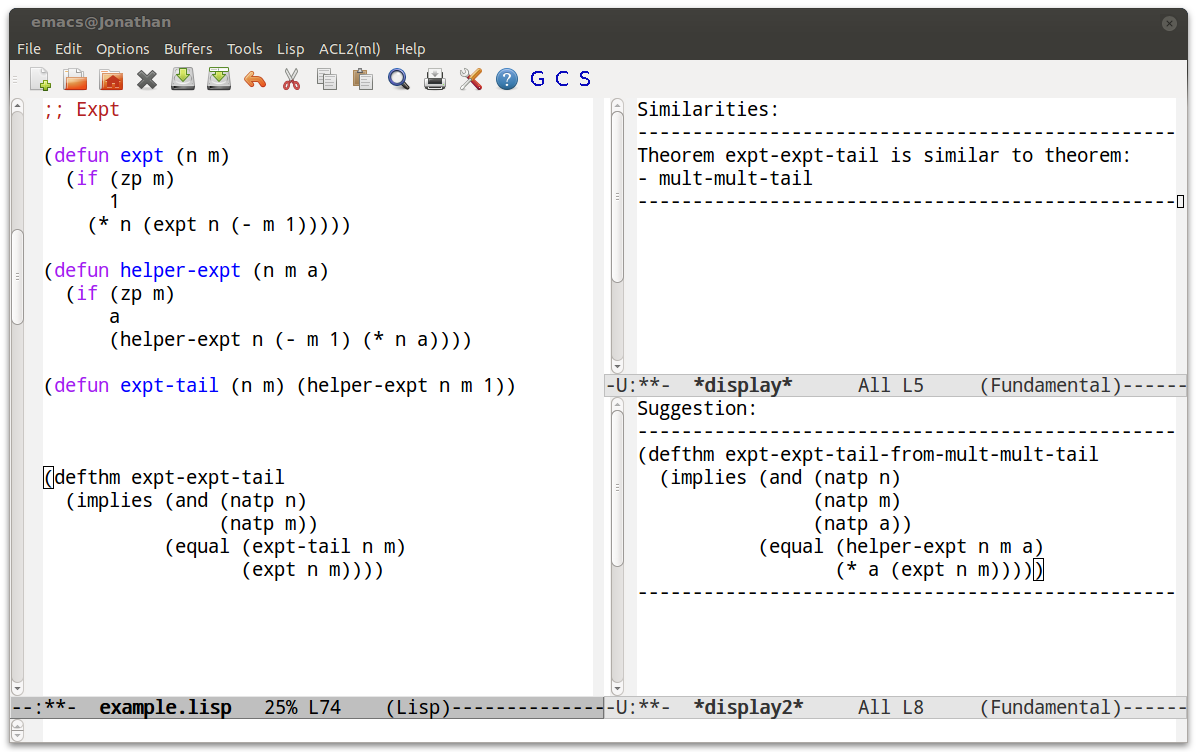}
\caption{{\scriptsize \emph{Suggestions for the lemma expt-expt-tail.} The Emacs window has been split into three buffers: the left buffer keeps the current ACL2 development, the right-top buffer shows the lemmas that are similar to expt-expt-tail in the current development, and the right-bottom buffer shows the suggestion generated to prove lemma expt-expt-tail.}}\label{fig:screenshot}
\end{figure}

\begin{example}\label{ex:multexptgen}
In Example~\ref{ex:multexpt}, for the case of exponentiation, we have: \verb"expt-expt-tail" is the TT, \verb"mult-mult-tail" is the ST, and \verb"mult-helper-mult" is the SL. ACL2(ml) uses this information to generate the following TL by mutating the SL:

\begin{verbatim}
(implies (and (natp n) (natp m) (natp a))
         (equal (helper-expt n m a) (* a (expt n m)))) 
\end{verbatim}

\noindent This is the necessary lemma to prove the TT --- Figure~\ref{fig:screenshot} shows how this feedback is presented in ACL2(ml). 
\end{example}

The lemma analogy tool implemented in ACL2(ml) consists of three levels of increasing term tree mutation. After each iteration, ACL2(ml) tests the validity of the set of generated equations using a counterexample checker. If no candidate conjecture survives counterexample checking, mutation proceeds to the next iteration. The first iteration, \emph{tree reconstruction}, replaces symbols in the SL with their analogical counterparts obtained from the TT. The second iteration, \emph{node expansion}, further mutates the term, by synthesising small terms in place of variables. Finally, the last iteration, \emph{term tree expansion}, similarly adds new term structure, but on the top-level of the term. The algorithms for these three iterations were presented in~\cite{lpar13}.

In~\cite{lpar13}, the lemma analogy tool was implemented as a standalone OCaml application. This tool was automatically invoked from ACL2(ml) when the user requested the generation of auxiliary lemmas. In the current version of ACL2(ml), the lemma analogy tool has been reimplemented in Emacs Lisp making the tool more efficient. 
This approach avoids the translation of terms to a different language and does not require the communication with an external tool. ACL2(ml) checks the generated conjectures using the counterexample generator implemented for ACL2 Sedan~\cite{testingACL2}.

We refer to the ACL2(ml) user manual~\cite{acl2ml} for a detailed description of how to use the tool.

\section{Clustering Functions of the ACL2 Community Books}\label{sec:clusterdef}


In this section, we present the first original contribution of this paper -- extrapolation of the theorem clustering described in~\cite{lpar13} to all ACL2 objects, including function definitions. In particular, we are interested in finding families of similar definitions; since the discovery of these families can avoid redundancies in the code and speed-up the proof-development.

Technically, we apply the same method of recurrent clustering, as described in the previous section, to all ACL2 definitions -- the difference is that ACL2(ml) clusters function definitions, instead of clustering theorem statements, as final step after the recurrent clustering cycle. 
%
We will use this section to show how this extension of ACL2(ml) can help to analyse redundancies in the ACL2 books repository.

ACL2 books accumulate models, definitions, proofs and proof libraries that can be applied in new proof developments. However, due to the size of the ACL2 books repository, it is a challenge to trace them and find reusable concepts and proof ideas. For example, there is a total of $23290$ definitions in the ACL2 distribution; then, it is unfeasible to manually detect patterns or redundancies arising in the function definitions. ACL2(ml) can perform this task automatically. Note that ACL2(ml) analyses definitions on the basis of their structure and their dependencies on other ACL2 functions; rather than on the concrete notation used in the definitions. Therefore, ACL2(ml) can detect cases when the same or similar notions are redefined with different names and using different notation.

Using its default configuration, ACL2(ml) discovers 981 clusters across the definitions of the ACL2 books repository in approximately $5$ minutes. From the 981 clusters, we can distinguish the following classification of clusters (see also~\cite{acl2ml} for a supporting extended note about this experiment).

\begin{itemize}

 \item[\textbf{C.1.}] $39\%$ of the clusters are related to a function that has been defined several times in different books of the same library. 
 
 \begin{example}
 The \verb"mergesort-exec" function is implemented in the \verb"sets" and \verb"sort" books of the \verb"coi" library. It is worth mentioning that in the definition of the \verb"sort" book, ACL2 needs two explicit hints to accept the definition of \verb"mergesort-exec"; on the contrary, in the \verb"sets" book, these two hints are not necessary. 
  \end{example}

 \item[\textbf{C.2.}] $18\%$ of the clusters are related to a function that has been re-defined in different versions of a library. 
 
 \begin{example}
 The \verb"rtl" library includes $5$ different versions, and several functions are repeated in all of them. For example, ACL2(ml) detects that the function \verb"logxor1" is defined in $4$ versions of this library.
 \end{example}

 \item[\textbf{C.3.}] $22\%$ of the clusters are related to a function that has been defined not only in different versions of a library, but also in different books of the same version of that library. 
 
 \begin{example}
Most of the clusters that belong to this case come from the \verb"rtl" library. As an example, ACL2(ml) finds the cluster containing the multiple definitions of \verb"fl", a function that has been defined $108$ times across the different versions and books of the \verb"rtl" library. 
 \end{example}

 \item[\textbf{C.4.}] $10\%$ of the clusters contain functions that have been defined in the same library and that are closely related. 
 
 \begin{example}
 An example of this kind of cluster is found in the \verb"centaur" library. ACL2(ml) discovers that the following two functions are in the same cluster.
 
\begin{verbatim}
(defun nonsubset-witness (x y)
  (declare (xargs :guard (and (setp x) (setp y))))
  (if (empty x)
      nil
    (if (in (head x) y)
        (nonsubset-witness (tail x) y)
      (head x))))

(defun intersectp-witness (x y)
  (declare (xargs :guard (and (setp x) (setp y))))
  (if (empty x)
      nil
    (if (in (head x) y)
        (head x)
      (intersectp-witness (tail x) y))))      
\end{verbatim}
The similarity between these two theorems can be easily spotted (both functions share the same structure and background functions), but note that this similarity is automatically discovered by ACL2(ml).

 \end{example}
 
 \item[\textbf{C.5.}] $5\%$ of the clusters are related to a function that has been defined several times in different libraries. These clusters are difficult to track manually since the same function is usually defined with different names; however, ACL2(ml) 
  finds these clusters based on the structure and background information of the function definitions.
 
 \begin{example}
 ACL2(ml) discovers that the libraries \verb"unicode" and \verb"security" define the same function to convert an ASCII character list into a Unicode string, but this function is called \verb"charlist=>ustring" in the \verb"unicode" library, and \verb"charlist-to-bytes" in the \verb"security" library. 
 \end{example}

  \item[\textbf{C.6.}] $6\%$ of the clusters consist of functions that are similar and are defined across several ACL2 libraries. 
 
 \begin{example}
The library \verb"coi" contains, in two different books, the following two close definitions that recognise respectively if all the elements of a list are not integers, and if a list contains an integer.

\begin{verbatim}
(defun all-not-integerp (x)
  (if (consp x)
      (if (integerp (car x))
          nil
        (all-not-integerp (cdr x)))
    t))
    
(defun some-integerp (x)
  (if (consp x)
      (if (integerp (car x))
          t
        (some-integerp (cdr x)))
    nil))    
\end{verbatim}
\end{example} 
 
\end{itemize}

In the above classification, ACL2(ml) discovers two kinds of clusters across ACL2 books: clusters related to a function that has been redefined in several books, and clusters of similar functions. The clusters that are classified as \textbf{C.1}--\textbf{C.3} could be found by searching function definitions with the same name. However, it is worth remarking that ACL2(ml)'s method goes beyond keyword searching and discovers functions that are defined with different names (clusters classified as \textbf{C.5}) and functions that are close related (clusters classified as \textbf{C.4} and \textbf{C.6}). In particular, the main properties that distinguish ACL2(ml) pattern search from the searching tools in ACL2 are:

\begin{itemize}
 \item the user does not have to know or provide any searching parameter;
 \item the discovered clusters do not have to follow a ``pattern'' in a strict sense (e.g. neither exact symbol names nor their strict order 
have to match exactly), but ACL2(ml) considers structures and background information found in the library; and,
 \item working with potentially huge sets of ACL2 objects, ACL2(ml) makes its own intelligent discrimination of more significant and less significant patterns, based on cluster statistics.
\end{itemize}

%
%
%
%

To conclude this section, it is worth mentioning that ACL2(ml) can also work in a goal-directed mode and discover only clusters of functions that are similar to a given function $f$. This can speed-up the proof development in two different ways. Clustering will provide definitions of functions similar to $f$; hence, the proofs of the theorems involving those functions may follow similar patterns, and may need similar auxiliary lemmas. Clustering can also discover that a newly defined function $f$ was previously defined; in that case, the user can use the existing library definition and all its background theory instead of defining it from scratch.

\section{A Precondition Generator based on ACL2 Guards}\label{sec:guards}

In this section, we present the second contribution of this paper -- an improvement of the lemma generation tool implemented in ACL2(ml). 
Given a target theorem, a source theorem, and a source lemma; the lemma analogy tool implemented in ACL2(ml) constructs a target lemma mutating the conclusion of the source lemma. Some of the mutated conjectures are true only under certain preconditions; however, these preconditions were not automatically generated in the initial ACL2(ml).


\begin{example}\label{ex:fact-fib}
Consider that we are proving the equivalence between the recursive and tail-recursive versions of the Fibonacci function (theorem \verb"fib-fib-tail" in Figure~\ref{fig:rec-tail}), and we request ACL2(ml) the generation of an auxiliary lemma for this target theorem. As a first step, ACL2(ml) finds the theorems that are similar to the theorem \verb"fib-fib-tail" --- as we have seen in Example~\ref{ex:multexpt}, this theorem is similar to the equivalence theorem about the factorial function (\verb"fact-fact-tail"). Subsequently, the lemma analogy tool will mutate the auxiliary lemma \verb"fact-helper-fact" (used in the proof of \verb"fact-fact-tail") trying to generate a target lemma.

Unfortunately, ACL2(ml) does not generate any valid suggestion -- all of them are falsified during the counterexample checking phase. However, inspecting the non-valid conjectures, we find the following result:
\verb"(equal (helper-fib n j k) (+ (* (fib (- n 1)) j) (* (fib n) k))". This is the rewriting rule that is necessary to complete the proof of \verb"fib-fib-tail", but the necessary preconditions are missed. Note that using directly the preconditions of \verb"fact-helper-fact" (as in Example~\ref{ex:multexptgen}) is not enough since we have an extra variable, and it is also necessary to impose the condition ``$n$ higher than $0$''.
\end{example}

%

We have addressed this problem using the \emph{guard} mechanism of ACL2~\cite{KMM00-1}. ACL2 is an untyped system; however, it provides a mechanism for restricting a function to a particular domain. Such  restrictions are called \emph{guards}, and they are used for two different aims: to increase the efficiency of the execution of ACL2 functions (proving that guards are satisfied allows ACL2 to directly use the underlying Common Lisp implementation to execute functions~\cite{efficientACL2}), or as a specification device to characterise the kinds of inputs a function should have (the ACL2 value universe is divided into 14 pairwise-disjoint ``primitive-types'' and the ACL2 user can create new ``types'' by defining a predicate that recognises a subset of the ACL2 universe). We are interested in the latter.

The use of guards as specification device resembles the use of types in typed-languages. Then, we can use this ``type'' information to generate preconditions for the conjectures generated by ACL2(ml). However, guards are not mandatory when the user defines a function, and several functions do not provide explicitly their guards (in these functions, \verb"T" is assigned as default guard value). To solve this problem, we have defined  Algorithm~\ref{alg:recurrent} to compute the guard on a function $f$ based on the guards of the functions called in the definition of $f$.

\begin{algorithm}
\KwIn{A function $f$.}
\KwOut{The guard $\bar{g}$ on $f$.}
\BlankLine
ask ACL2 the stored guard $\bar{g}$ on $f$\;
\eIf{$\bar{g}\neq T$}
   {\Return{$\bar{g}$.}}
   {$\bar{g}\leftarrow$ \texttt{T}\;
    expand the macros in the definition of $f$\;
   \For{\textnormal{\textbf{each}} function call $(f_i~~t_i^1~\ldots ~t_i^{n_i})$ included in the expanded definition of $f$}{
      \eIf{guard $g_i$ on $f_i$ was memoised $(f_i,g_i)$}
          {$\bar{g}_i \leftarrow g_i$ where its parameters are replaced with $t_i^1~\ldots ~t_i^{n_i}$\;
           $\bar{g}\leftarrow (\bar{g} $ and $\bar{g}_i)$;}
   {ask ACL2 the stored guard $g_i$ on the function $f_i$\;
   \Case{$g_i \neq$ \texttt{T}:}{memoise $(f_i,g_i)$ for further use\;
          $\bar{g}_i \leftarrow g_i$ where its parameters are replaced with $t_i^1~\ldots ~t_i^{n_i}$\;
          $\bar{g}\leftarrow (\bar{g} $ and $\bar{g}_i)$\;}
   \Case{$g_i =$ \texttt{T} $and$ either $f_i$ is an ACL2 built-in function or $f_i=f$:}{memoise $(f_i,\texttt{T})$ for further use\;}
   \Other{compute guard $g_i$ on $f_i$ using Algorithm~\ref{alg:recurrent}\;
          memoise $(f_i,g_i)$ for further use\;
          $\bar{g}_i \leftarrow g_i$ where its parameters are replaced with $t_i^1~\ldots ~t_i^{n_i}$\;
          $\bar{g}\leftarrow (\bar{g} $ and $\bar{g}_i)$;}
    \Return{simplification of $\bar{g}$.}}}
   }
\caption{Recurrent Guard Generation}
\label{alg:recurrent}
\end{algorithm}

ACL2(ml) uses Algorithm~\ref{alg:recurrent} to compute the guard on a function, after it is defined in ACL2; in addition, the computed guard is stored for further use. 
The implementation of Algorithm~\ref{alg:recurrent} in ACL2(ml) invokes ACL2 with two different aims:

\begin{itemize}
 \item obtain the guard information stored in ACL2: we use the keyword command \verb":args" to see the guard on a function; and,
 \item simplification of the final result: we use the \verb"easy-simplify-term" function included in the book ``tools/easy-simplify.lisp'' to simplify the guard $\bar{g}$.
\end{itemize}

%
%

\begin{example}
Let us use Algorithm~\ref{alg:recurrent} to obtain the guards of the function \verb"helper-fib" (cf. Figure~\ref{fig:rec-tail}). This function calls the functions \verb"zp", \verb"equal", \verb"binary-+" (\verb"-" and \verb"+" are macros that are expanded to the function \verb"binary-+") and \verb"helper-fib" (note that \verb"helper-fib" is a recursive function). The guards of these functions are given in Table~\ref{tab:guards}. After combining and simplifying these guards, the result is: \verb"(and (integerp n) (not (< n 0)) (acl2-numberp j)" \verb"(acl2-numberp k))". 

\begin{table}
\centering 
{\tiny
\begin{tabular}{|c|c|c|c|}
 \hline
 \rowcolor{black!20!white}\texttt{Function call} & \texttt{Function with formals} & \texttt{Guard} & \texttt{Guard with arguments}\\
 \hline
 \texttt{(zp n)}& \texttt{(zp x)}& \texttt{(and (integerp x) (<= x 0))}& \texttt{(and (integerp n) (<= n 0))}\\
 \hline
 \texttt{(equal n 1)} & \texttt{(equal x y)} & \texttt{T} &\texttt{T}\\
 \hline
  \texttt{(binary-+ n -1)} & \texttt{(binary-+ x y)} &\texttt{(and (acl2-numberp x) (acl2-numberp y))} &\texttt{(and (acl2-numberp n) (acl2-numberp -1))}\\
 \hline
  \texttt{(binary-+ j k)} &   \texttt{(binary-+ x y)} & \texttt{(and (acl2-numberp x) (acl2-numberp y))} &\texttt{(and (acl2-numberp j) (acl2-numberp k))}\\
 \hline
 \texttt{(helper-fib (- n 1) k (+ j k))} & \texttt{(helper-fib n j k)} &\texttt{T} & \texttt{T} \\
\hline
\end{tabular}}
\caption{{\scriptsize \emph{Guards of the functions used in the definition of \texttt{helper-fib}.}}}\label{tab:guards}
\end{table}

\end{example}

Given a conjecture $c$ constructed by the lemma analogy tool of ACL2(ml), the guards of the functions invoked in $c$ can be used to generate the preconditions of $c$, see Algorithm~\ref{alg2}.

\begin{algorithm}
\KwIn{The conclusion of a conjecture $c$.}
\KwOut{Precondition $p$ of $c$ generated from the guards of the functions used in $c$.}
\BlankLine
$p\leftarrow$ \texttt{T}\;
expand the macros in the conjecture $c$\;
\For{\textnormal{\textbf{each}} function call $(f_i~~t_i^1~\ldots ~t_i^{n_i})$ included in the expanded conjecture $c$}{
use Algorithm~\ref{alg:recurrent} to compute the guard $g_i$ on $f_i$\;
$\bar{g}_i \leftarrow g_i$ where its parameters are replaced with $t_i^1~\ldots ~t_i^{n_i}$\;
$p\leftarrow (p $ and $\bar{g}_i)$\;
}
\Return{simplification of $p$.}
\caption{Generation of Preconditions}
\label{alg2}
\end{algorithm}

\begin{example}\label{ex:conjecture}
Let us see the preconditions that are generated using Algorithm~\ref{alg2} for the conjecture \verb"(equal (helper-fib n j k) (+ (* (fib (- n 1)) j) (* (fib n) k)))" from Example~\ref{ex:fact-fib}. The guards for the different functions included in this conjecture are given in Table~\ref{tab:guards2}. After combining and simplifying the guards, the generated precondition is: 
\begin{verbatim}
(and (integerp n) (not (< n 0)) 
     (acl2-numberp j) (acl2-numberp k) 
     (not (< (+ -1 n) 0)))
\end{verbatim}

\noindent These are the necessary conditions to prove the given conjecture -- provided that the arithmetic library is included in the development.
\end{example}

It is worth mentioning that the preconditions generated by ACL2(ml) go beyond type information (for instance, in Example~\ref{ex:conjecture} the generated preconditions impose that \verb"n" must be a natural number higher than $0$). Note also that the preconditions generated by the guards may not be enough to prove a conjecture, since the necessary preconditions could be more complex than the information included in the guards.

The generator of preconditions is internally called by the lemma analogy tool of ACL2(ml); additionally, the user can also invoke it to obtain the preconditions of his own conjectures. This functionality can be especially useful for novice users of ACL2, since they often assume that a conjecture is restricted to the domain of interest when specifying their conjectures. However, the ACL2 logic needs all assumptions to be given explicitly; hence, the generator of preconditions can help them to add the missing preconditions to their conjectures. Another application of this tool is the discovery of false conjectures: if the guards generated from the functions of a conjecture contradict themselves, 
the user is trying to prove a result that is either false or meaningless. 

\begin{example}
Consider the tail-recursive function to compute the length of a list:

\begin{verbatim}
(defun lengthTail (lst res)
   (if (endp lst) res (lengthTail (cdr lst) (1+ res))))
\end{verbatim}

If we try to prove the conjecture \verb"(equal (+ res (length lst)) (lengthTail res lst))" (where we have swapped the arguments of the function \verb"lengthTail"); and invoke the generator of preconditions, the returned result is \verb"nil". This indicates that there is a contradiction in the guards of the functions: the guard generated from \verb"+" is \verb"(acl2-numberp res)", and the guard generated from \verb"lengthTail" is \verb"(consp res)" -- this means that \verb"res" should belong to two disjoint ``types''; hence, the conjecture is false. If we fix the conjecture putting the arguments in the correct position, the generator of preconditions returns the necessary assumption, \verb"(acl2-numberp res)", to prove the lemma.

\end{example}

\begin{table}
\centering 
{\tiny
\begin{tabular}{|l|l|l|l|}
 \hline
 \rowcolor{black!20!white}\texttt{Function call}& \texttt{Function with formals} & \texttt{Guards} & \texttt{Guards with arguments}\\
 \hline
 \texttt{(fib n)}&  \texttt{(fib x)}& \texttt{(and (integerp x) (<= x 0))}& \texttt{(and (integerp n) (<= n 0))}\\
 \hline
 \texttt{(binary-* (fib n) k)} & \texttt{(binary-* x y)} &  \texttt{(and (acl2-numberp x) (acl2-numberp y))} &\texttt{(and (acl2-numberp (fib n))}\\
   &  &   &\texttt{~~~~~(acl2-numberp k))}\\
 \hline
  \texttt{(binary-+ n -1)} & \texttt{(binary-+ x y)}& \texttt{(and (acl2-numberp x) (acl2-numberp y))} &\texttt{(and (acl2-numberp n) (acl2-numberp -1))}\\
 \hline
  \texttt{(fib (- n 1))} &  \texttt{(fib x)} &\texttt{(and (integerp x) (<= x 0))} &\texttt{(and (integerp (- n 1)) (<= (- n 1) 0))}\\
 \hline
\texttt{(binary-* (fib (- n 1)) j)}& \texttt{(binary-+ x y)} & \texttt{(and (acl2-numberp x) (acl2-numberp y))} &\texttt{(and (acl2-numberp (fib (- n 1)))}\\
 &  &   &\texttt{~~~~~(acl2-numberp j))}\\
  \hline
\texttt{(binary-+ (* (fib (- n 1)) j)}& \texttt{(binary-+ x y)} & \texttt{(and (acl2-numberp x) (acl2-numberp y))} &\texttt{(and (acl2-numberp (* (fib (- n 1)) j)) }\\
\texttt{~~~~~~~~~~(* (fib n) k))}& && \texttt{~~~~~(acl2-numberp (* (fib n) k)))}\\
  \hline
\texttt{(helper-fib n j k)} & \texttt{(helper-fib n j k)} &\texttt{(and (integerp n) (not (< n 0))} & \texttt{(and (integerp n) (not (< n 0))} \\
& &\texttt{~~~~~(acl2-numberp j) (acl2-numberp k))}& \texttt{~~~~~(acl2-numberp j) (acl2-numberp k))}\\
\hline
\end{tabular}}
\caption{{\scriptsize \emph{Guards of the functions used in the conjecture \texttt{(equal (helper-fib n j k) (+ (* (fib (- n 1) j)) (* (fib n) k))}.}}}\label{tab:guards2}
\end{table}

\section{Related Work, Conclusions and Further Work}\label{sec:conclusions}

\paragraph{Related Work.} ACL2(ml) combines statistical and symbolic machine-learning methods to guide the ACL2 user during the proof development. Statistical machine-learning methods can discover data regularities based on numeric proof features. Among other successful statistical tools is the method of statistical proof-premise selection~\cite{DFGS99,DenzingerS00,KuhlweinLTUH12,TsivtsivadzeUGH11,UrbanSPV08}. Applied in several automated (first-order) theorem provers, this method provides statistical ratings of existing lemmas; and this information is used to make automated rewriting more efficient. There are several differences between the premise selection tools and ACL2(ml):

\begin{itemize}
 \item Both the premise selection tools and ACL2(ml) use matrices to encode term trees, but the former uses sparse incidence matrices (capturing the occurrence of the symbols of a library in the term tree) and the latter uses a special feature matrix (cf. Figure~\ref{tree}) that captures the structure and dependencies of the function symbols occurring in the tree.  
 \item The premise selection tools use supervised machine-learning techniques -- in particular, a classifier is constructed for every lemma $L$ of a given library, and such a classifier is trained to estimate the likelihood of using the lemma $L$ during a proof. On the contrary, ACL2(ml) uses unsupervised clustering to discover families of similar theorems and definitions. 
 \item The aim of the premise selection tools is to make automated rewriting more efficient -- the methods applied in these tools could be used to improve the efficiency of ACL2's rewriter. Then, these tools are focused on helping the prover; on the contrary, ACL2(ml) was born as a hint provider to help the user in the discovery of similarities across libraries and the formulation of auxiliary results. 
\end{itemize}

The task of formulating auxiliary lemmas can be undertaken either manually or using symbolic machine learning techniques. While statistical methods focus on extracting information from existing large theory libraries, symbolic methods are instead concerned with automating the discovery of lemmas in new  theories~\cite{HR,mathsaid,JDB11,montano2012,hipspec,HLW12}, while relying on existing proof strategies, e.g. proof-planning and rippling \cite{BB05}. These methods can be slow on large inputs due to the increase in the search space. This problem is overcome in ACL2(ml) feeding the output of statistical machine-learning algorithms (families of similar theorems) to a mutation tool that uses this data for more efficient lemma discovery.

\paragraph{Conclusions and Further Work.} In this paper, we have presented two extensions for ACL2(ml) that advance ACL2(ml) further in the direction of guiding the ACL2 user during the proof development. 
The first extension allows the user to detect families of similar function definitions. This can be used to reduce the redundancies in ACL2 books, reuse previous developments, and find similar definitions that can be helpful for the user's current development. The second extension enhances the quality of the conjectures constructed by ACL2(ml), not only analogising the conclusions of theorems but also generating some preconditions using the guard mechanism of ACL2. This extension can also be invoked directly by the user to generate preconditions for his own conjectures. 

ACL2(ml) is currently implemented as an extension for Emacs (one of the two main ACL2 development environments), we are planning to reimplement some of ACL2(ml) modules (namely, the generators of conjectures and preconditions) as ACL2 books to integrate them in other environments, or use them without running Emacs. We are also exploring different alternatives to improve the heuristics to generate auxiliary lemmas (for instance, using the information shown by ACL2 in failed proof-attempts), and studying how to generate new functions based on the statistical results obtained using clustering. In addition, we are interested in enhancing the generator of preconditions. In general, ACL2 lemmas should be stated as general as possible keeping the number of hypothesis to a minimum; however, the generator of preconditions can introduce unnecessary hypothesis in the conjectures constructed by ACL2(ml). The \verb"remove-hyps" tool~\cite{WK14} was introduced to remove unnecessary hypothesis from theorems; and therefore 
it would be helpful to incorporate this tool into ACL2(ml). 

\section*{Acknowledgements}

We are grateful to the anonymous referees for their useful suggestions and comments; and J Moore for useful discussions about this work.

\bibliographystyle{eptcs}
\bibliography{biblio}

\begin{thebibliography}{10}
\providecommand{\bibitemdeclare}[2]{}
\providecommand{\surnamestart}{}
\providecommand{\surnameend}{}
\providecommand{\urlprefix}{Available at }
\providecommand{\url}[1]{\texttt{#1}}
\providecommand{\href}[2]{\texttt{#2}}
\providecommand{\urlalt}[2]{\href{#1}{#2}}
\providecommand{\doi}[1]{doi:\urlalt{http://dx.doi.org/#1}{#1}}
\providecommand{\bibinfo}[2]{#2}

\bibitemdeclare{book}{BB05}
\bibitem{BB05}
\bibinfo{author}{D.~\surnamestart Basin\surnameend} et~al.
  (\bibinfo{year}{2005}): \emph{\bibinfo{title}{Rippling: Meta-level Guidance
  for Mathematical Reasoning}}.
\newblock \bibinfo{publisher}{Cambridge University Press}.

\bibitemdeclare{book}{Bishop}
\bibitem{Bishop}
\bibinfo{author}{C.~\surnamestart Bishop\surnameend} (\bibinfo{year}{2006}):
  \emph{\bibinfo{title}{Pattern Recognition and Machine Learning}}.
\newblock \bibinfo{publisher}{Springer}.

\bibitemdeclare{inproceedings}{Brock96}
\bibitem{Brock96}
\bibinfo{author}{B.~\surnamestart Brock\surnameend},
  \bibinfo{author}{M.~\surnamestart Kaufmann\surnameend} \&
  \bibinfo{author}{{J}~S. \surnamestart Moore\surnameend}
  (\bibinfo{year}{1996}): \emph{\bibinfo{title}{ACL2 Theorems about Commercial
  Microprocessors}}.
\newblock In: {\sl \bibinfo{booktitle}{1st International Conference on Formal
  Methods in Computer-Aided Design (FMCAD'96)}}, {\sl \bibinfo{series}{LNCS}}
  \bibinfo{volume}{1166}, pp. \bibinfo{pages}{275--293},
  \doi{10.1007/BFb0031816}.

\bibitemdeclare{inproceedings}{testingACL2}
\bibitem{testingACL2}
\bibinfo{author}{H.~\surnamestart Chamarthi\surnameend},
  \bibinfo{author}{P.~\surnamestart Dillinger\surnameend},
  \bibinfo{author}{M.~\surnamestart Kaufmann\surnameend} \&
  \bibinfo{author}{P.~\surnamestart Manolios\surnameend}
  (\bibinfo{year}{2011}): \emph{\bibinfo{title}{Integrating Testing and
  Interactive Theorem Proving}}.
\newblock In: {\sl \bibinfo{booktitle}{10th International Workshop on the ACL2
  Theorem Prover and its Applications (ACL2'11)}}, {\sl
  \bibinfo{series}{EPTCS}}~\bibinfo{volume}{70}, pp. \bibinfo{pages}{4--19},
  \doi{10.4204/EPTCS.70.1}.

\bibitemdeclare{inproceedings}{hipspec}
\bibitem{hipspec}
\bibinfo{author}{K.~\surnamestart Claessen\surnameend} et~al.
  (\bibinfo{year}{2013}): \emph{\bibinfo{title}{Automating Inductive Proofs
  using Theory Exploration}}.
\newblock In: {\sl \bibinfo{booktitle}{24th International Conference on
  Automated Deduction (CADE-24)}}, {\sl \bibinfo{series}{LNCS}}
  \bibinfo{volume}{7898}, pp. \bibinfo{pages}{392--406},
  \doi{10.1007/978-3-642-38574-2\_27}.

\bibitemdeclare{inproceedings}{HR}
\bibitem{HR}
\bibinfo{author}{S.~\surnamestart Colton\surnameend} (\bibinfo{year}{2002}):
  \emph{\bibinfo{title}{{The HR Program for Theorem Generation}}}.
\newblock In: {\sl \bibinfo{booktitle}{18th International Conference on
  Automated Deduction (CADE-18)}}, {\sl \bibinfo{series}{LNCS}}
  \bibinfo{volume}{2392}, pp. \bibinfo{pages}{285--289},
  \doi{10.1007/3-540-45620-1\_24}.

\bibitemdeclare{misc}{acl2books}
\bibitem{acl2books}
\bibinfo{author}{J.~\surnamestart Davis\surnameend} et~al.:
  \emph{\bibinfo{title}{{ACL2 Community Books}}}.
\newblock \bibinfo{note}{\url{https://code.google.com/p/acl2-books/}}.

\bibitemdeclare{misc}{xdocacl2}
\bibitem{xdocacl2}
\bibinfo{author}{J.~\surnamestart Davis\surnameend} et~al.
  (\bibinfo{year}{2009--2014}): \emph{\bibinfo{title}{{XDOC Documentation
  System for ACL2}}}.
\newblock
  \bibinfo{note}{\url{http://www.cs.utexas.edu/users/moore/acl2/manuals/curren%
t/manual/}}.

\bibitemdeclare{techreport}{DFGS99}
\bibitem{DFGS99}
\bibinfo{author}{J.~\surnamestart Denzinger\surnameend},
  \bibinfo{author}{M.~\surnamestart Fuchs\surnameend},
  \bibinfo{author}{C.~\surnamestart Goller\surnameend} \&
  \bibinfo{author}{S.~\surnamestart Schulz\surnameend} (\bibinfo{year}{1999}):
  \emph{\bibinfo{title}{Learning from Previous Proof Experience: A Survey}}.
\newblock \bibinfo{type}{Technical Report}, \bibinfo{institution}{Technische
  Universitat Munchen}.

\bibitemdeclare{article}{DenzingerS00}
\bibitem{DenzingerS00}
\bibinfo{author}{J.~\surnamestart Denzinger\surnameend} \&
  \bibinfo{author}{S.~\surnamestart Schulz\surnameend} (\bibinfo{year}{2000}):
  \emph{\bibinfo{title}{Automatic Acquisition of Search Control Knowledge from
  Multiple Proof Attempts}}.
\newblock {\sl \bibinfo{journal}{Information and Computation}}
  \bibinfo{volume}{162}(\bibinfo{number}{1-2}), pp. \bibinfo{pages}{59--79},
  \doi{10.1006/inco.1999.2857}.

\bibitemdeclare{inproceedings}{EVF07}
\bibitem{EVF07}
\bibinfo{author}{C.~\surnamestart Eastlund\surnameend},
  \bibinfo{author}{D.~\surnamestart Vaillancourt\surnameend} \&
  \bibinfo{author}{M.~\surnamestart Felleisen\surnameend}
  (\bibinfo{year}{2007}): \emph{\bibinfo{title}{{ACL2 for Freshmen: First
  Experiences}}}.
\newblock In: {\sl \bibinfo{booktitle}{7th International Workshop on the ACL2
  Theorem Prover and its Applications (ACL2'07)}}, \bibinfo{series}{ACM Press},
  pp. \bibinfo{pages}{200--211}.

\bibitemdeclare{article}{efficientACL2}
\bibitem{efficientACL2}
\bibinfo{author}{D.~\surnamestart Greve\surnameend} et~al.
  (\bibinfo{year}{2008}): \emph{\bibinfo{title}{Efficient Execution in an
  Automated Reasoning Environment}}.
\newblock {\sl \bibinfo{journal}{Journal of Functional Programming}}
  \bibinfo{volume}{18}(\bibinfo{number}{1}), pp. \bibinfo{pages}{15--46},
  \doi{10.1017/S0956796807006338}.

\bibitemdeclare{article}{Weka}
\bibitem{Weka}
\bibinfo{author}{M.~\surnamestart Hall\surnameend} et~al.
  (\bibinfo{year}{2009}): \emph{\bibinfo{title}{{The WEKA Data Mining Software:
  An Update}}}.
\newblock {\sl \bibinfo{journal}{SIGKDD Explorations}}
  \bibinfo{volume}{11}(\bibinfo{number}{1}), pp. \bibinfo{pages}{10--18},
  \doi{10.1145/1656274.1656278}.

\bibitemdeclare{misc}{acl2ml}
\bibitem{acl2ml}
\bibinfo{author}{J.~\surnamestart Heras\surnameend} \&
  \bibinfo{author}{E.~\surnamestart Komendantskaya\surnameend}
  (\bibinfo{year}{2013--2014}): \emph{\bibinfo{title}{{ACL2(ml): downloadable
  programs, manual, examples}}}.
\newblock
  \bibinfo{note}{\url{http://staff.computing.dundee.ac.uk/katya/ACL2ml}}.

\bibitemdeclare{inproceedings}{lpar13}
\bibitem{lpar13}
\bibinfo{author}{J.~\surnamestart Heras\surnameend},
  \bibinfo{author}{E.~\surnamestart Komendantskaya\surnameend},
  \bibinfo{author}{M.~\surnamestart Johansson\surnameend} \&
  \bibinfo{author}{E.~\surnamestart Maclean\surnameend} (\bibinfo{year}{2013}):
  \emph{\bibinfo{title}{{Proof-Pattern Recognition and Lemma Discovery in
  ACL2}}}.
\newblock In: {\sl \bibinfo{booktitle}{19th Logic for Programming Artificial
  Intelligence and Reasoning (LPAR-19)}}, {\sl \bibinfo{series}{LNCS}}
  \bibinfo{volume}{8312}, pp. \bibinfo{pages}{389--406},
  \doi{10.1007/978-3-642-45221-5\_27}.

\bibitemdeclare{inproceedings}{HLW12}
\bibitem{HLW12}
\bibinfo{author}{S.~\surnamestart Hetzl\surnameend},
  \bibinfo{author}{A.~\surnamestart Leitsch\surnameend} \&
  \bibinfo{author}{D.~\surnamestart Weller\surnameend} (\bibinfo{year}{2012}):
  \emph{\bibinfo{title}{Towards Algorithm Cut-Introduction}}.
\newblock In: {\sl \bibinfo{booktitle}{18th Logic for Programming Artificial
  Intelligence and Reasoning (LPAR-18)}}, {\sl \bibinfo{series}{LNCS}}
  \bibinfo{volume}{7180}, pp. \bibinfo{pages}{228--242},
  \doi{10.1007/978-3-642-28717-6\_19}.

\bibitemdeclare{article}{JDB11}
\bibitem{JDB11}
\bibinfo{author}{M.~\surnamestart Johansson\surnameend},
  \bibinfo{author}{L.~\surnamestart Dixon\surnameend} \&
  \bibinfo{author}{A.~\surnamestart Bundy\surnameend} (\bibinfo{year}{2011}):
  \emph{\bibinfo{title}{Conjecture Synthesis for Inductive Theories}}.
\newblock {\sl \bibinfo{journal}{Journal of Automated Reasoning}}
  \bibinfo{volume}{47}(\bibinfo{number}{3}), pp. \bibinfo{pages}{251--289},
  \doi{10.1007/s10817-010-9193-y}.

\bibitemdeclare{book}{KMM00}
\bibitem{KMM00}
\bibinfo{editor}{M.~\surnamestart Kaufmann\surnameend},
  \bibinfo{editor}{P.~\surnamestart Manolios\surnameend} \&
  \bibinfo{editor}{\surnamestart {J S. Moore}\surnameend}, editors
  (\bibinfo{year}{2000}): \emph{\bibinfo{title}{Computer-Aided Reasoning: ACL2
  Case Studies}}.
\newblock \bibinfo{publisher}{Kluwer Academic Publishers},
  \doi{10.1007/978-1-4757-3188-0}.

\bibitemdeclare{book}{KMM00-1}
\bibitem{KMM00-1}
\bibinfo{editor}{M.~\surnamestart Kaufmann\surnameend},
  \bibinfo{editor}{P.~\surnamestart Manolios\surnameend} \&
  \bibinfo{editor}{{J}~S. \surnamestart Moore\surnameend}, editors
  (\bibinfo{year}{2000}): \emph{\bibinfo{title}{Computer-Aided Reasoning: An
  approach}}.
\newblock \bibinfo{publisher}{Kluwer Academic Publishers},
  \doi{10.1007/978-1-4615-4449-4}.

\bibitemdeclare{inproceedings}{KuhlweinLTUH12}
\bibitem{KuhlweinLTUH12}
\bibinfo{author}{D.~\surnamestart K{\"u}hlwein\surnameend},
  \bibinfo{author}{T.~\surnamestart van Laarhoven\surnameend},
  \bibinfo{author}{E.~\surnamestart Tsivtsivadze\surnameend},
  \bibinfo{author}{J.~\surnamestart Urban\surnameend} \&
  \bibinfo{author}{T.~\surnamestart Heskes\surnameend} (\bibinfo{year}{2012}):
  \emph{\bibinfo{title}{Overview and Evaluation of Premise Selection Techniques
  for Large Theory Mathematics}}.
\newblock In: {\sl \bibinfo{booktitle}{6th International Joint Conference on
  Automated Reasoning (IJCAR'12)}}, {\sl \bibinfo{series}{LNCS}}
  \bibinfo{volume}{7364}, pp. \bibinfo{pages}{378--392},
  \doi{10.1007/978-3-642-31365-3\_30}.

\bibitemdeclare{inproceedings}{mathsaid}
\bibitem{mathsaid}
\bibinfo{author}{R.~L. \surnamestart McCasland\surnameend},
  \bibinfo{author}{A.~\surnamestart Bundy\surnameend} \& \bibinfo{author}{P.~F.
  \surnamestart Smith\surnameend} (\bibinfo{year}{2006}):
  \emph{\bibinfo{title}{Ascertaining Mathematical Theorems}}.
\newblock In: {\sl \bibinfo{booktitle}{12th Symposium on the Integration of
  Symbolic Computation and Mechanized Reasoning (Calculemus 2005)}}, {\sl
  \bibinfo{series}{ENTCS}} \bibinfo{volume}{151}, pp. \bibinfo{pages}{21--38},
  \doi{10.1016/j.entcs.2005.11.021}.

\bibitemdeclare{article}{montano2012}
\bibitem{montano2012}
\bibinfo{author}{O.~\surnamestart Montano-Rivas\surnameend},
  \bibinfo{author}{R.~\surnamestart Mccasland\surnameend},
  \bibinfo{author}{L.~\surnamestart Dixon\surnameend} \&
  \bibinfo{author}{A.~\surnamestart Bundy\surnameend} (\bibinfo{year}{2012}):
  \emph{\bibinfo{title}{Scheme-based theorem discovery and concept invention}}.
\newblock {\sl \bibinfo{journal}{Expert Systems with Applications}}
  \bibinfo{volume}{39}(\bibinfo{number}{2}), pp. \bibinfo{pages}{1637--1646},
  \doi{10.1016/j.eswa.2011.06.055}.

\bibitemdeclare{article}{JFP:1381480}
\bibitem{JFP:1381480}
\bibinfo{author}{R.~\surnamestart Page\surnameend} (\bibinfo{year}{2007}):
  \emph{\bibinfo{title}{Engineering Software Correctness}}.
\newblock {\sl \bibinfo{journal}{Journal of Functional Programming}}
  \bibinfo{volume}{17}(\bibinfo{number}{6}), pp. \bibinfo{pages}{675--686},
  \doi{10.1017/S095679680700634X}.

\bibitemdeclare{inproceedings}{TsivtsivadzeUGH11}
\bibitem{TsivtsivadzeUGH11}
\bibinfo{author}{E.~\surnamestart Tsivtsivadze\surnameend},
  \bibinfo{author}{J.~\surnamestart Urban\surnameend},
  \bibinfo{author}{H.~\surnamestart Geuvers\surnameend} \&
  \bibinfo{author}{T.~\surnamestart Heskes\surnameend} (\bibinfo{year}{2011}):
  \emph{\bibinfo{title}{Semantic Graph Kernels for Automated Reasoning}}.
\newblock In: {\sl \bibinfo{booktitle}{SIAM Conference on Data Mining
  (SDM'11)}}, \bibinfo{publisher}{SIAM / Omnipress}, pp.
  \bibinfo{pages}{795--803}, \doi{10.1137/1.9781611972818.68}.

\bibitemdeclare{inproceedings}{UrbanSPV08}
\bibitem{UrbanSPV08}
\bibinfo{author}{J.~\surnamestart Urban\surnameend},
  \bibinfo{author}{G.~\surnamestart Sutcliffe\surnameend},
  \bibinfo{author}{P.~\surnamestart Pudl{\'a}k\surnameend} \&
  \bibinfo{author}{J.~\surnamestart Vyskocil\surnameend}
  (\bibinfo{year}{2008}): \emph{\bibinfo{title}{{MaLARea SG1- Machine Learner
  for Automated Reasoning with Semantic Guidance}}}.
\newblock In: {\sl \bibinfo{booktitle}{4th International Joint Conference on
  Automated Reasoning (IJCAR'08)}}, {\sl \bibinfo{series}{LNCS}}
  \bibinfo{volume}{5195}, pp. \bibinfo{pages}{441--456},
  \doi{10.1007/978-3-540-71070-7\_37}.

\bibitemdeclare{misc}{WK14}
\bibitem{WK14}
\bibinfo{author}{N.~\surnamestart Wetzler\surnameend} \&
  \bibinfo{author}{M.~\surnamestart Kaufmann\surnameend}
  (\bibinfo{year}{2014}): \emph{\bibinfo{title}{{Remove-hyps and Writing
  Utilities with Make-event}}}.
\newblock \bibinfo{note}{ACL2 Theorem Proving Seminar. University of Texas}.

\end{thebibliography}
\end{document}